\documentclass[fleqn,twoside,psfig]{article}
\usepackage{espcrc2}
\usepackage{refmerge} 
\usepackage{psfig}
\usepackage[figuresright]{rotating}
\input definitions
\input myBabarsym

\def\ggg{g}
\title{The Hadronic Contribution to \boldmath$(\ggg-2)_\mu$}

\author{Michel Davier\address[LAL]{Laboratoire de l'Acc\'el\'erateur 
				   Lin\'eaire, 
				IN2P3/CNRS-Universit\'e Paris-Sud 11, \\ 
        		           BP34, 91898 Orsay, France}\\
 		                   E-mail: davier@lal.in2p3.fr}

\begin{document}
\begin{abstract}
The evaluation of the hadronic contribution to the muon magnetic 
anomaly $a_\mu$ is revisited, taking advantage of new experimental
data on $e^+e^-$ annihilation into hadrons: SND and CMD-2 for the 
$\pi^+\pi^-$ channel, and \babar for multihadron final states.
Discrepancies are observed between KLOE and CMD-2/SND data, preventing
one from averaging all the $e^+e^-$ results. The long-standing disagreement
between spectral functions obtained from $\tau$ decays and $e^+e^-$
annihilation is still present, and not accounted by 
isospin-breaking corrections, for which new estimates have been 
presented. The updated Standard Model value for $a_\mu$ based on $e^+e^-$
annihilation data is now reaching a precision better than experiment,
and it disagrees with the direct measurement from BNL at the 3.3$\sigma$ 
level, while the $\tau$-based estimate is in much better agreement.
The $\tau$/$e^+e^-$ discrepancy, best revealed when comparing the
measured branching fraction for $\tau^- \to \pi^- \pi^0 \nu_\tau$ to
its prediction from the isospin-breaking-corrected $e^+e^-$ spectral 
function, remains a serious problem to be understood.
\end{abstract}

\maketitle
%
%

\section{Introduction}

Hadronic vacuum polarization (HVP) in the photon propagator plays an 
important role in many precision tests of the Standard Model. 
This is the case for the muon magnetic anomaly
$a_\mu\equiv(g_\mu -2)/2$, where the HVP component is the leading 
contributor to the uncertainty of the Standard Model prediction.
The HVP contribution is computed by means of a dispersion relation
as an integral over experimentally determined spectral functions.
It is a property of this dispersion relation that the $\pi\pi$ 
spectral function provides the major part of the total HVP
contribution, so that the experimental effort focuses on this channel.

Spectral functions are directly obtained from the  
measured cross sections of \epem annihilation into hadrons. 
The accuracy of the HVP predictions has therefore followed 
the progress in the quality of the data~\cite{eidelman} it relies on.
Because the data quality was not always suitable, it was deemed necessary to 
resort to other sources of information. One such possibility was the 
use of the vector spectral functions~\cite{adh} derived from the study 
of hadronic $\tau$ decays~\cite{aleph_vsf} for the energy range less 
than $m_\tau\simeq1.8\gevcc$. For this purpose, 
the isospin rotation that leads from 
the charged $\tau$ to the neutral \epem  final state has to be 
thoroughly corrected for isospin-breaking effects.

Also, it was demonstrated that essentially perturbative QCD could be 
applied to energy scales as low as $1$--$2\gev$~\cite{aleph_asf,opal_alphas},
thus offering a way to replace poor \epem data in some energy regions 
by a reliable and precise theoretical 
prescription~\cite{martin,dh97,steinhauser,erler,groote,dh98}. 

Detailed reanalyses including all available experimental data have been
published in Refs.~\cite{dehz03,dehz,teubner,yndurain,jegerlehner}, 
taking advantage of precise results in the $\pi\pi$ channel from the CMD-2 
experiment~\cite{cmd2_new} and from the ALEPH analysis of $\tau$ 
decays~\cite{aleph_new}, and benefiting from a more complete
treatment of isospin-breaking corrections~\cite{ecker1,ecker2}.
With the increased accuracy of the \epem data a discrepancy with
the isospin-breaking-corrected $\tau$ spectral functions was 
found~\cite{dehz03}, thus leading to inconsistent predictions for the 
lowest-order hadronic contribution to $a_\mu$. The dominant 
contribution to this discrepancy stems from the $\pi^+\pi^-$ channel,
although another discrepancy occurs in the $\pi^+\pi^-2\pi^0$ mode.

Improvements in the HVP calculation are needed in order to match the
present experimental accuracy on $a_\mu$
from the BNL experiment~\cite{bnl_2006},
\beq
\label{eq:bnlexp}
	a_\mu \:=\: (11\,659\,208.0 \pm 6.3)\tmten~.
\eeq

In this paper I revisit the input to the HVP dispersion integral in the
light of new experimental data on $e^+e^- \to \pi^+ \pi^-$ from
SND~\cite{snd_corr} and CMD-2~\cite{cmd2_high,cmd2_low,cmd2_rho},
and on multihadron final states from 
\babar~\cite{babar_3pi,babar_4pi,babar_6pi,wenfeng} 
using the radiative return technique~\cite{isr}. These new measurements 
represent a significant step forward, as they overcome in precision
previous determinations in the same channels.

%
%

\section{Muon Magnetic Anomaly}
\label{sec:anomaly}

It is convenient to separate the Standard Model (SM) prediction for the
anomalous magnetic moment of the muon
into its different contributions,
\beq
    a_\mu^{\rm SM} \:=\: a_\mu^{\rm QED} + a_\mu^{\rm had} +
                             a_\mu^{\rm weak}~,
\eeq
with
\beq
 a_\mu^{\rm had} \:=\: a_\mu^{\rm had,LO} + a_\mu^{\rm had,HO}
           + a_\mu^{\rm had,LBL}~,
\eeq
and where $a_\mu^{\rm QED}=(11\,658\,471.9\pm0.1)\tmten$ is 
the pure electromagnetic contribution~\cite{kino-nio}, 
\amuhadLO\ is the lowest-order HVP contribution, 
$a_\mu^{\rm had,HO}=(-9.8\pm0.1)\tmten$ 
is the corresponding higher-order part~\cite{krause2,adh,teubner}, 
and $a_\mu^{\rm weak}=(15.4\pm0.1\pm0.2)\tmten$,
where the first error is the hadronic uncertainty and the second
is due to the Higgs mass range, accounts for corrections due to
exchange of the weakly interacting bosons up to two loops~\cite{amuweak}. 
For the light-by-light (LBL) scattering part, $a_\mu^{\rm had,LBL}$,
we use the value $(12.0\pm3.5)\tmten$ from the latest 
evaluation~\cite{lbl_mv,dm}, slightly corrected for the missing 
contribution from (mainly) the pion box.

Owing to unitarity and to the analyticity of the vacuum-polarization 
function, the lowest order HVP contribution to $a_\mu$ can be computed 
\via\ the dispersion integral~\cite{rafael}
\beq
\label{eq_int_amu}
    a_\mu^{\rm had,LO} \:=\: 
           \frac{\alpha^2(0)}{3\pi^2}
           \intl_{4m_\pi^2}^\infty\!\!ds\,\frac{K(s)}{s}R^{(0)}(s)~,
\eeq
where $K(s)$ is a well-known QED kernel, and
$R^{(0)}(s)$ denotes the ratio of the ``bare'' cross
section for \epem annihilation into hadrons to the pointlike muon-pair cross
section. The bare cross section is defined as the measured cross section
corrected for initial-state radiation, electron-vertex loop contributions
and vacuum-polarization effects in the photon propagator. However, photon 
radiation in the final state is included in the bare cross section 
defined here. The reason for using the bare (\ie, lowest order) 
cross section is that a full treatment of higher orders is anyhow 
needed at the level of $a_\mu$, so that the use of the ``dressed''
cross section would entail the risk of double-counting some of the 
higher-order contributions.

The function $K(s)\sim1/s$ in Eq.~(\ref{eq_int_amu}) gives a strong 
weight to the low-energy part of the integral. About 91\pc\ of the 
total contribution to \amuhadLO\  is accumulated at center-of-mass
energies $\sqrt{s}$ below $1.8\gev$ and 73\pc\ of \amuhadLO\ is covered 
by the $\pi\pi$ final state, which is dominated by the $\rho(770)$ 
resonance. 

%
%
\section{The $2\pi$ Input Data}

A detailed compilation of all the experimental data used in the evaluation 
of the dispersion integral~(\ref{eq_int_amu}) prior to 2004 is provided 
in Refs.~\cite{dehz,dehz03}. Also discussed therein is the corrective 
treatment of radiative effects applied to some of the measurements.
The $\tau$ $2\pi$ spectral function is obtained by averaging the results 
from ALEPH~\cite{aleph_vsf}, CLEO~\cite{cleo_2pi} and OPAL~\cite{opal_2pi},
which exhibit satisfactory mutual agreement. Since 2004, new 
cross section measurements became available
from KLOE~\cite{kloe_pipi} using radiative return at DAPHNE and
from the annihilation experiments SND~\cite{snd_corr} and 
CMD-2~\cite{cmd2_high,cmd2_low,cmd2_rho} at Novosibirsk.

\begin{figure*}[t]  
  \centerline{
	  \psfig{file=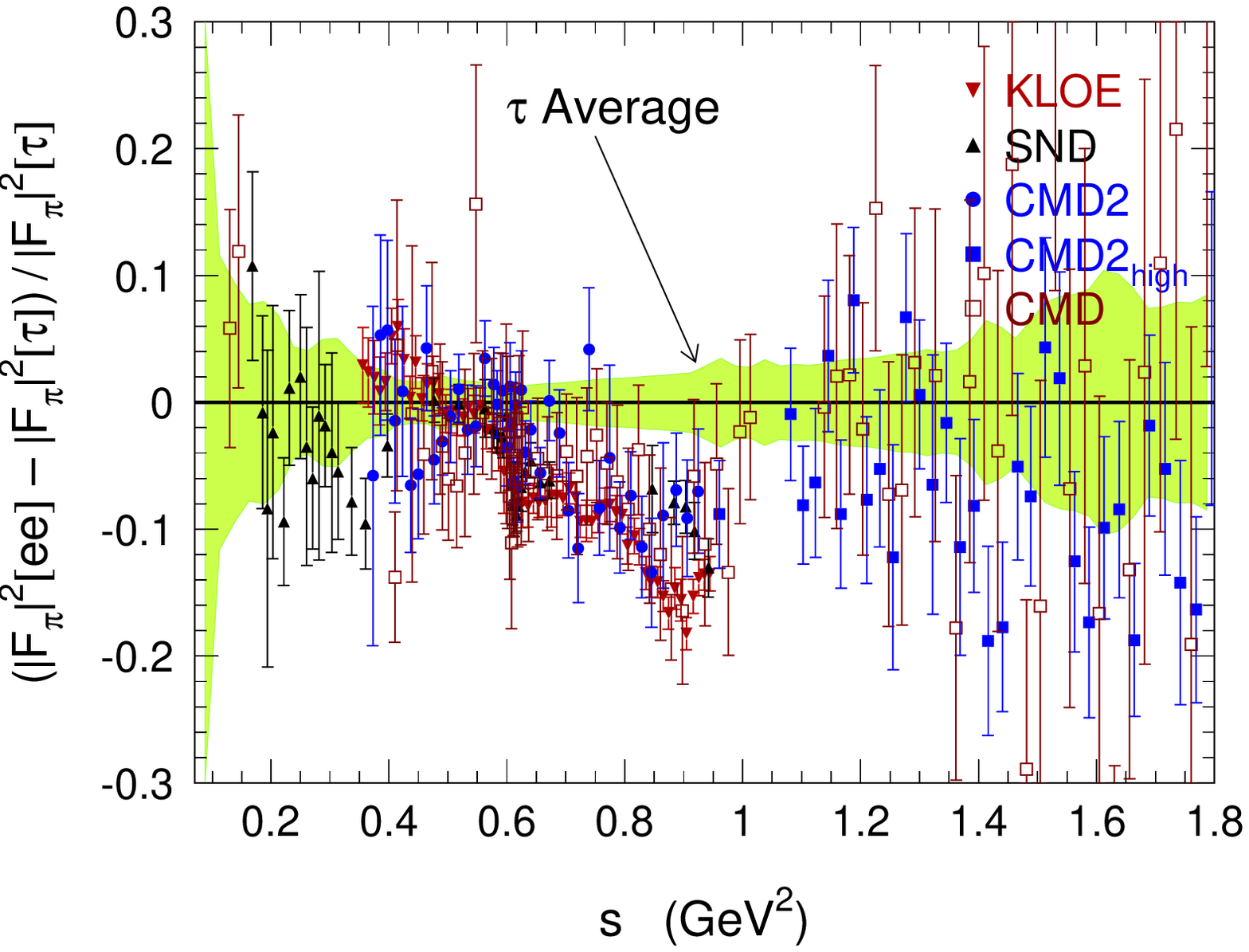,width=80mm}
          \hspace{0.05cm}
	  \psfig{file=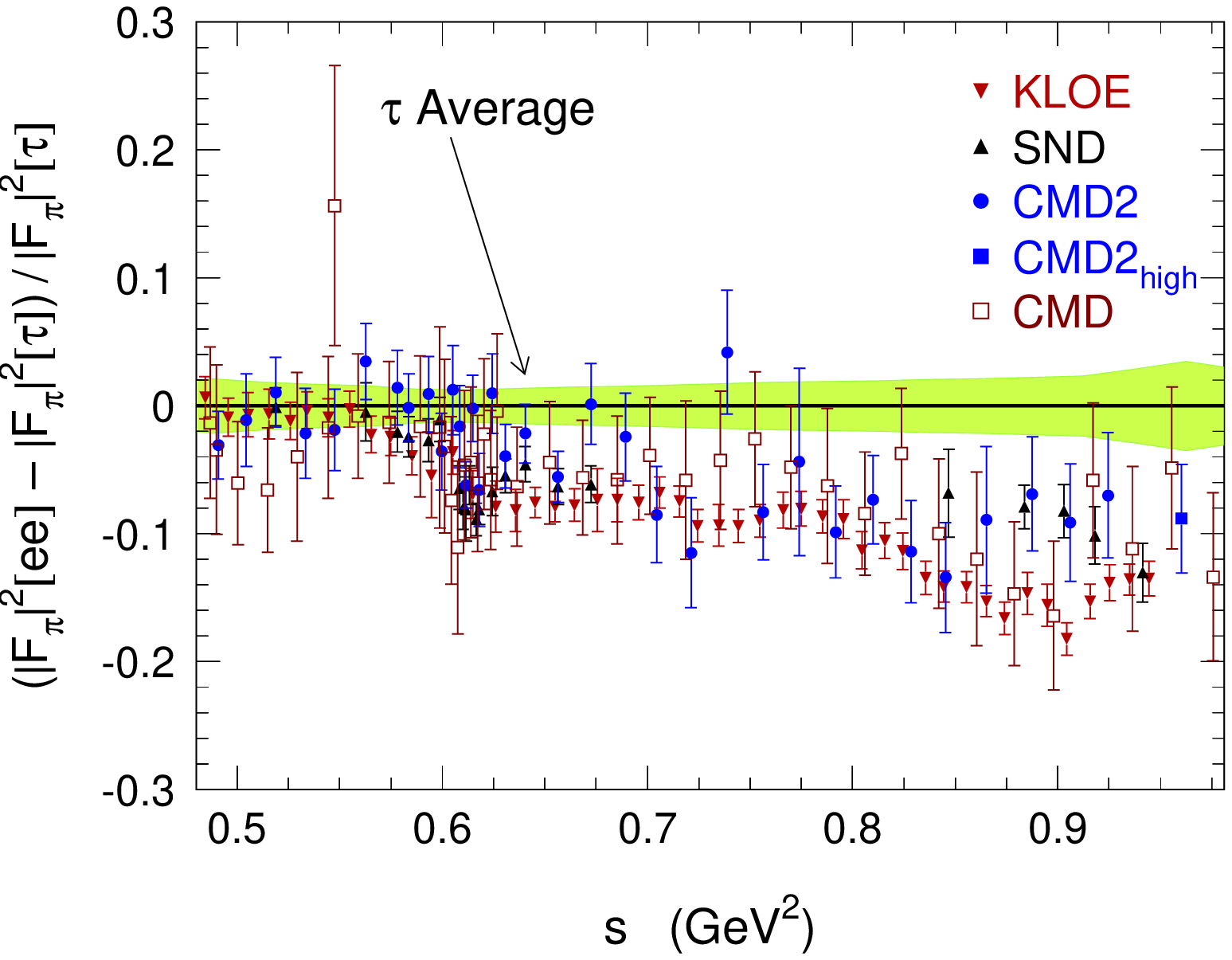,width=80mm}
	     }
  \vspace{-1cm}
  \caption[.]{\label{fig:2pi_comp} \it
	Relative comparison of the $\pi^+\pi^-$ spectral functions
    	from \epem-annihilation data and isospin-breaking-corrected 
	$\tau$ data, expressed as a ratio to the $\tau$ spectral function.
	The shaded band indicates the errors of the $\tau$ data.
        The \epem  data are from KLOE~\cite{kloe_pipi}, 
	CMD-2~\cite{cmd2_new}, CMD, OLYA and DM1 
        (references given in Ref.~\cite{dehz03}).
        The right hand plot emphasizes the region of the $\rho$ 
        peak.}  
\end{figure*}

The comparison between all new $e^+e^-$ results and the combined $\tau$
spectral function is given in Fig.~\ref{fig:2pi_comp}.

A few remarks are in order: 
\begin{itemize}
\item Revision of the radiative corrections applied the CMD-2 (94-95 data)
and SND data led to corrections amounting up to 3\%.
\item The revised SND and CMD-2 (also including new data released in 2006)
spectral functions agree within errors. It should be pointed out that
both analyses now use the same radiative correction package, introducing a
full correlation between the two data sets.
\item The high-statistics KLOE data do not agree with SND and CMD-2, mostly
through a discrepancy in the $\rho$ lineshape: KLOE is higher below the
peak and becomes lower above.
\item A significant discrepancy, most visible above the $\rho$ peak,
but present almost everywhere, is found between $\tau$ and the \epem data.  
\end{itemize}

Considering different $\rho^-$ and $\rho^0$ masses as an additional 
isospin-breaking correction of the $\tau$ spectral function  would improve 
the comparison in the $\rho$ resonance peak, but in that case the discrepancy
should fade away for masses above $m_\rho + \Gamma_\rho /2$, which is 
not observed.

Finally, some improvement of the isospin-breaking corrections has been 
proposed in Ref.~\cite{castro}. A contribution from the $\rho\omega\pi$
vertex with $\omega \to \pi^0 \gamma$ was not included in the treatment
used so far~\cite{ecker2}, as it occurs from higher order in Chiral
Perturbation Theory. It was however found to be significant, but its 
effect amounts to only 20\% of the observed discrepancy.

\section{Testing CVC}

Measurement of branching fractions in $\tau$ decays are more robust than
the spectral functions, as the latter ones depend on the experimental 
resolution and require a numerically delicate unfolding. 
It is possible to relate the measured
branching ratios for $\tau^- \to V^- \nu_\tau$, where $V$ is any vector
final state, to their expectations from CVC using $e^+e^-$ spectral
functions, duly corrected for isospin breaking. In this way we do not
anymore rely on the shape of the $\tau$ spectral function and 
instead concentrate on the relative normalization and the isospin-breaking 
corrections.

\begin{figure}
  \centerline{        
	  \psfig{file=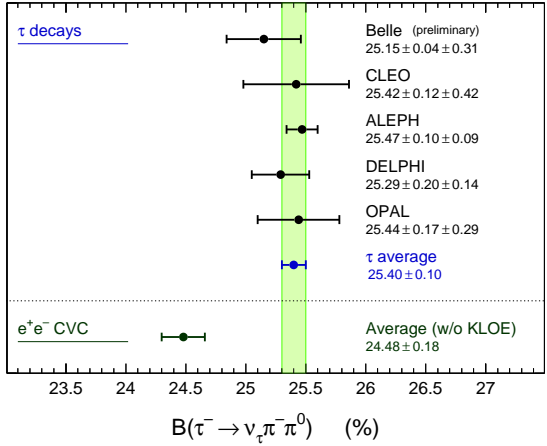,width=75mm}
             }
  \vspace{-0.4cm}
\caption[.]{\it The measured branching ratios for 
       $\tau^-\rightarrow\nu_\tau\pi^-\pi^0$ compared to the prediction
       from the $e^+e^-\rar\pi^+\pi^-$ spectral function applying the 
       isospin-breaking correction factors discussed in Ref.~\cite{dehz}.
       The measured branching ratios are from ALEPH~\cite{aleph_new},
       CLEO~\cite{cleo_bpipi0} and OPAL~\cite{opal_bpipi0}.
       The L3 and OPAL results are obtained from their $h \pi^0$ branching 
       ratio, reduced by the small $K \pi^0$ contribution 
       measured by ALEPH~\cite{aleph_ksum} and CLEO~\cite{cleo_kpi0}.}
\label{brpipi0}
\end{figure}

The result of the test for the $\pi^- \pi^0$ channel is shown in 
Fig.~\ref{brpipi0}. It shows a large discrepancy between the average $\tau$
branching ratio and the CVC prediction. The difference
$[B_\tau-B_{CVC}]_{\pi \pi^0}= (0.92 \pm 0.21)$\% is 4.5$\sigma$ away from
zero. In relative terms, the discrepancy is a 3.6\% effect, about twice the
already applied isospin-breaking correction, dominated by 
(expected to be) well-controlled short-distance effects.

The same exercise can be applied to other channels. For the $\pi 3\pi^0$
final state, the difference is consistent with zero, $-(0.08 \pm 0.11)$\%.
The result for the $3\pi \pi^0$ mode is bad, $(0.91 \pm 0.25)$\%, 
amounting to a relative 20\% discrepancy, certainly much beyond any
reasonable isospin-breaking effect. Here the quality of the data on 
$e^+e^- \to \pi^+ \pi^- 2\pi^0$ is in doubt, as they show a considerable
spread between the experiments.

\section{New Multihadron Data}

Results from the \babar~\cite{babar_3pi,babar_4pi,babar_6pi,wenfeng} 
experiment are being produced on the multihadron
final states using radiative return~\cite{isr}. They are part 
of a program designed to cover most exclusive annihilation processes in the
few GeV energy range, taking advantage of the large initial centre-of-mass
energy of 10.6 GeV. Hard-radiated photons are detected at large angle,
together with the hadronic system byproducts, so that the full final state
can be kinematically constrained (see the discussion in
Ref.~\cite{wenfeng}). These results significantly improve the
corresponding contributions to $a_\mu^{\rm had}$, 
since earlier data in the 1.4-3 GeV
energy range were of poor statistical and systematic quality.

\begin{figure}
  \centerline{
	  \psfig{file=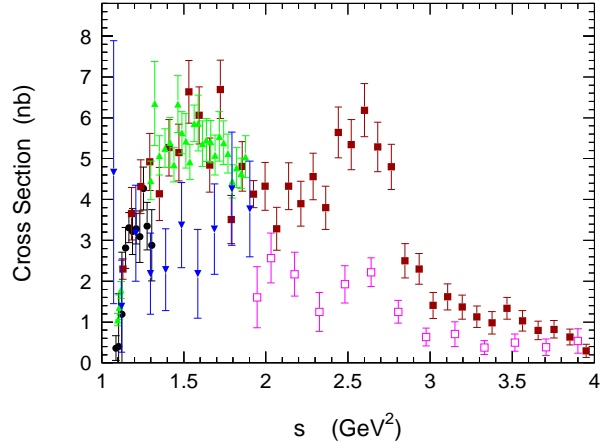,width=80mm}
             }
  \vspace{-1cm}
\caption[.]{\it The measured cross section for $e^+e^- \to \pi^+\pi^-\pi^0$ 
       from \babar~\cite{babar_3pi} (indicated by full squares) compared to 
       previous measurements (see references in Ref.~\cite{dehz}).}
\label{2pipi0}
\end{figure}

\begin{figure}
  \centerline{
	  \psfig{file=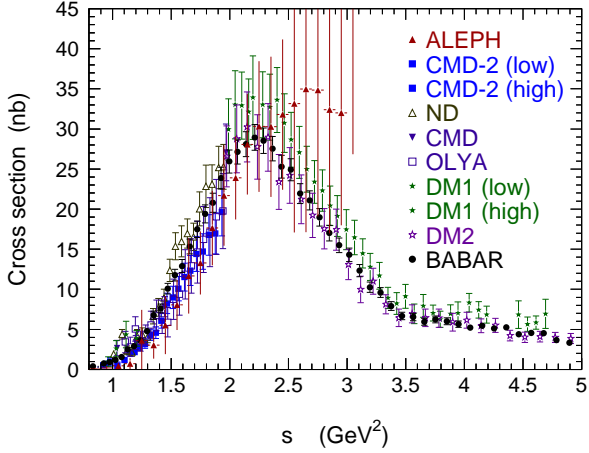,width=80mm}
             }
  \vspace{-1cm}
\caption[.]{\it The measured cross section for $e^+e^- \to 2\pi^+2\pi^-$ 
       from \babar~\cite{babar_4pi} (indicated by full circles) compared to 
       previous measurements 
       and results from $\tau^- \to \pi^- 3\pi^0$ (see references 
       in Ref.~\cite{dehz}).}
\label{4pi}
\end{figure}

\begin{figure}
  \centerline{
	  \psfig{file=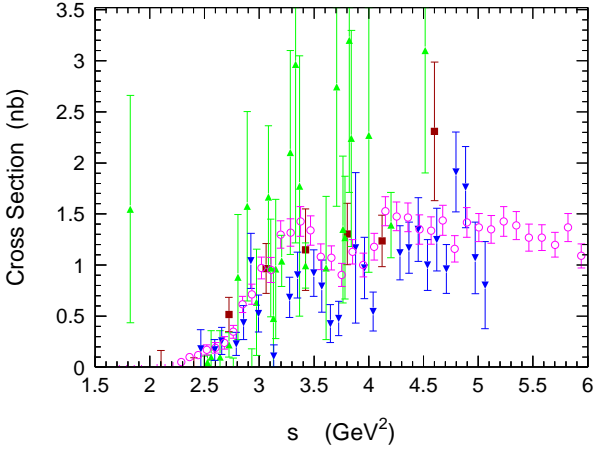,width=80mm}
             }
  \vspace{-1cm}
\caption[.]{\it The measured cross section for $e^+e^- \to 3\pi^+3\pi^-$ 
       from \babar~\cite{babar_6pi} (indicated by open circles) compared to 
       previous measurements (see references in Ref.~\cite{dehz}).}
\label{6pi}
\end{figure}

\begin{figure}
  \centerline{
	  \psfig{file=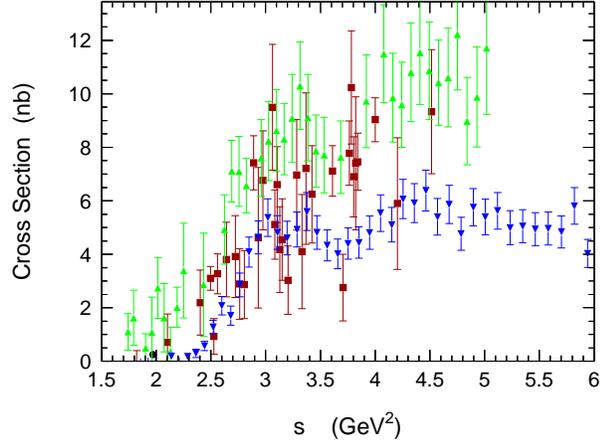,width=80mm}
             }
  \vspace{-1cm}
\caption[.]{\it The measured cross section 
       for $e^+e^- \to 2\pi^+2\pi^-2\pi^0$ from \babar~\cite{babar_6pi} 
       (indicated by triangles pointing below) compared to previous 
       measurements (see references in Ref.~\cite{dehz}).}
\label{4pi2pi0}
\end{figure}

The \babar\ results are compared to other data  in 
Figs.~\ref{2pipi0}-\ref{4pi2pi0} for
the $e^+e^- \to \pi^+\pi^-\pi^0$, $2\pi^+2\pi^-$, $3\pi^+3\pi^-$, and
$2\pi^+2\pi^-2\pi^0$ cross sections. Besides a generally good agreement
with previous experiments, some important differences are seen for
$\pi^+\pi^-\pi^0$ and the 6-pion states with the results obtained at DCI.
The impact of these new measurements is quantified in Table~\ref{babar-amu}.

\begin{table}
\caption{\it The contribution of some multipion processes to 
$a_\mu^{\rm had}$ integrated from threshold to 1.8 GeV for the older 
experiments (references in Ref.~\cite{dehz}) and including the new \babar\ 
ISR results~\cite{babar_3pi,babar_4pi,babar_6pi,wenfeng}. 
Values are given in units of $10^{-10}$.}
  \vspace{0.5cm}
  \begin{center}
\begin{tabular}
{l|c|c} 
\hline\noalign{\smallskip}
   process   & older exp. & with \babar\  \\
\noalign{\smallskip}\hline\noalign{\smallskip}
 $\pi^+\pi^-\pi^0$    & $2.45 \pm 0.26$     & $3.25 \pm 0.09$  \\
 $2\pi^+2\pi^-$       & $14.20 \pm 0.90$    & $13.09 \pm 0.44$ \\
 $3\pi^+3\pi^-$       & $0.10 \pm 0.10$     & $0.11 \pm 0.02$  \\
 $2\pi^+2\pi^-2\pi^0$ & $1.42 \pm 0.30$     & $0.89 \pm 0.09$  \\
\noalign{\smallskip}\hline\noalign{\smallskip} 
\noalign{\smallskip}\hline
  \end{tabular}
\label{babar-amu}
  \end{center}
\end{table}

%
%
\section{Results}
\label{sec_results_amu}

During the previous evaluations of \amuhadLO, the results using 
respectively the $\tau$ and \epem  data were quoted individually, 
but on the same footing since the \epem-based evaluation was 
dominated by the data from a single experiment (CMD-2).
The confirmation of the $\tau$/$e^+e^-$ discrepancy by SND and KLOE 
may suggest to prefer the \epem-based result
until a better understanding of the dynamical origin of the 
observed effect is achieved. This discrepancy is a challenging problem,
which may itself turn out to be of fundamental importance. The present
update is therefore only based on $e^+e^-$ data, dominated by the latest
results from CMD-2 and SND which are in good agreement. The KLOE data
show a systematic trend which is not explained at the moment, so we
do not use them as the resulting increase in precision would not be
trustworthy. Further studies by KLOE are ongoing, in particular
a determination of the $\pi\pi/\mu\mu$ ratio where several systematic
effects cancel, which should lead a significant improvement of the
systematic uncertainty~\cite{kloe_tau06}.

The preliminary estimate of the integral~(\ref{eq_int_amu}) 
given below includes one 
additional improvement with respect to Ref.~\cite{dehz03}:
perturbative QCD is used instead of experimental data in the region 
between $1.8$ and $3.7\gev$, where non-perturbative contributions to 
integrals over
differently weighed spectral functions were found to be small~\cite{dh97}.
This results in a reduction of \amuhadLO \  by $-1\tmten$.
All contributions to the dispersion integral where no new input data
are available are taken from Ref.~\cite{dehz03}.

\begin{figure*} [t]
\centerline{\psfig{file=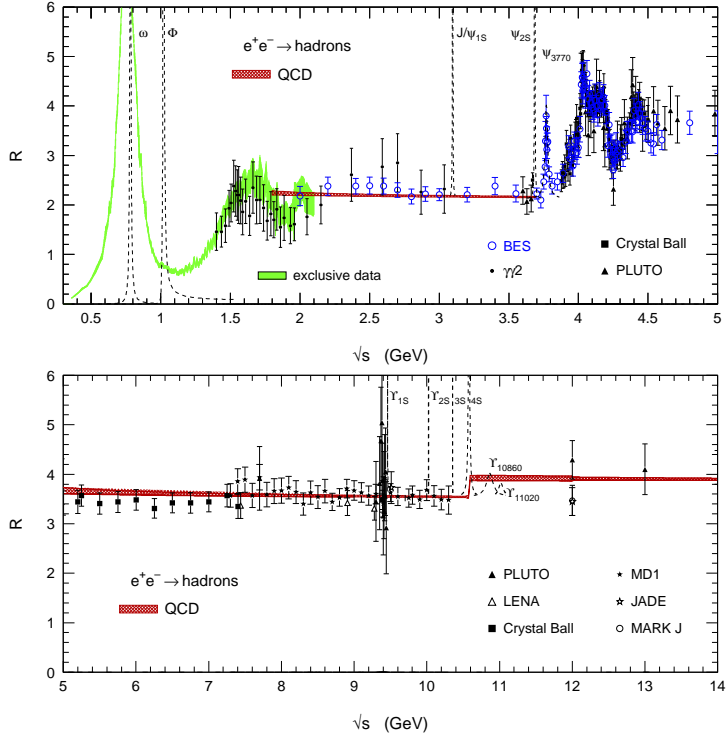,width=100mm}}
  \vspace{-0.4cm}
\caption[.]{\label{R_plot} \it
	The $R$ values from data and QCD used in the dispersion integral.
        The shaded band below 2.1 GeV represents the sum of the exclusive 
        channels, not yet including the new \babar\ results. Narrow resonances
        are indicated as dashed lines. All data points shown correspond to 
        inclusive measurements. The cross-hatched band gives the prediction
        from QCD used in the calculation from 1.8 to 3.7 GeV and above 5 GeV.
	}
\end{figure*}

The $R$ values from data and QCD are displayed in Fig.~\ref{R_plot}, but
not yet updated with the \babar\ multipion data. For masses larger than 
1.8 GeV, except in the $c\overline{c}$ threshold region from 3.7 to 5 GeV, 
the QCD prediction is used. Agreement between QCD and data is good.
The contributions of the different exclusive channels and of the continuum
are given in Table~\ref{contrib}.

The \epem-based result for the lowest order hadronic contribution is
\beq
  \amuhadLO =
       (690.8 \pm 3.9 \pm 1.9_{\rm rad} \pm 0.7_{\rm QCD})\tmten ~,
\label{amueehad}
\eeq
where the second error is due to our treatment of (potentially) 
missing radiative corrections in the older data~\cite{dehz}. For comparison,
the $\tau$-based result~\cite{dehz03} can be updated using the new \epem 
\babar data for the channels other than $2\pi$ or $4\pi$ yielding the value 
$(710.3 \pm 5.2)\tmten $. 
Adding to the \epem result~(\ref{amueehad}) the QED, higher-order hadronic, 
light-by-light scattering, and weak contributions given in 
Section~\ref{sec:anomaly},
one finds
\beqn
\label{eq:smres}
  a_{\mu}^{\rm SM} &=& (11\,659\,180.5
                     \pm 4.4_{\rm had,LO+HO}~~~~~~~~~ \nonumber\\
		   && \hspace{-0.8cm}
                     \pm\, 3.5_{\rm had,LBL} 
                     \pm 0.2_{\rm QED+EW})\tmten~.
\eeqn

\begin{figure} [t]
\centerline{\psfig{file=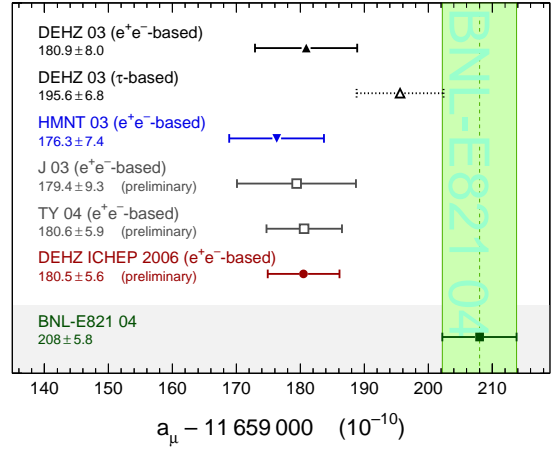,width=75mm}}
  \vspace{-0.cm}
\caption[.]{\label{fig:results} \it
	Comparison of the result~(\ref{eq:smres}) with the 
	BNL measurement~\cite{bnl_2006}. Also given
	are our previous estimates~\cite{dehz03}, where the 
	triangle with the dotted error bar indicates the 
	$\tau$-based result, as well as the estimates from 
	Refs.~\cite{teubner,yndurain,jegerlehner}, not yet including 
	the KLOE data.
	}
\end{figure}

This value can be compared to the present measurement~(\ref{eq:bnlexp});
adding all errors in quadrature, the difference between experiment
and theory is
\beq
\label{eq:diffbnltheo}
  a_\mu^{\rm exp}-a_{\mu}^{\rm SM} =
	(27.5\pm8.4)\tmten~,
\eeq
which corresponds to 3.3 ``standard deviations'' (to be interpreted
with care due to the dominance of systematic errors in the SM
prediction).
A graphical comparison of the result~(\ref{eq:smres}) with 
previous evaluations~\footnote{Results similar to ours have just appeared
in Ref.~\cite{HMNT06}.} and the experimental value is given in 
Fig.~\ref{fig:results}. 

\begin{table*} [t]
\caption{\it The contributions of different final states in specified energy
ranges to $a_\mu^{\rm had}$, given in units of $10^{-10}$. Major improvements
since Ref.~\cite{dehz03} are obtained for (1) the $\pi^+\pi^-$ channel from
CMD-2 (here preliminary results are used; final data have been 
published~\cite{cmd2_high,cmd2_low,cmd2_rho} since)
and SND~\cite{snd_corr}, while KLOE
results are not included (see text), and (2) the $2\pi^+2\pi^-$ and other
exclusive channels from \babar~\cite{babar_3pi,babar_4pi,babar_6pi,wenfeng}.
The uncertainty for missing radiative corrections, labeled 'rad', only 
concerns our ad hoc treatment~\cite{dehz} applied to older experimental 
data.}
  \vspace{0.5cm}
  \begin{center}
\setlength{\tabcolsep}{0.0pc}
\begin{tabular*}
{\textwidth}{@{\extracolsep{\fill}}lcc}
\hline\noalign{\smallskip}
  Modes & Energy range (GeV) & $a_\mu^{\rm had}$  \\
\noalign{\smallskip}\hline\noalign{\smallskip}
$\pi^+\pi^-$        &  $2m_\pi$--0.5    &  $55.6 \pm 0.8 \pm 0.1_{\rm rad}$  \\
$\pi^+\pi^-$        &  0.5--1.8         & $449.0 \pm 3.0 \pm 0.9_{\rm rad}$  \\
$2\pi^+2\pi^-$      &  $2m_\pi$--1.8    &  $13.1 \pm 0.4 \pm 0.0_{\rm rad}$  \\
$\pi^+\pi^-2\pi^0$  &  $2m_\pi$--1.8    &  $16.8 \pm 1.3 \pm 0.2_{\rm rad}$  \\
$\omega$            &  0.3--0.81        &  $38.0 \pm 1.0 \pm 0.3_{\rm rad}$  \\
$\phi$              &  1.0--1.055       &  $35.7 \pm 0.8 \pm 0.2_{\rm rad}$  \\
other exclusive         &  $2m_\pi$--1.8    &  $24.3 \pm 1.3 \pm 0.2_{\rm rad}$  \\
$J/\psi,\psi({2S})$ &  --               &   $7.4 \pm 0.4 \pm 0.0_{\rm rad}$  \\
R(QCD)              &  1.8--3.7         &  $33.9 \pm 0.5_{\rm QCD}$  \\
R(data)             &  3.7--5.0         &   $7.2 \pm 0.3 \pm 0.0_{\rm rad}$  \\
R(QCD)              &  5.0--$\infty$    &   $9.9 \pm 0.2_{\rm QCD}$  \\
\noalign{\smallskip}\hline\noalign{\smallskip}
sum                 & $2m_\pi$--$\infty$& ~~~~~~~~$690.8\pm3.9\pm1.9_{\rm rad}\pm0.7_{\rm QCD}$~~~~~~~~ \\
\noalign{\smallskip}\hline\noalign{\smallskip}
\noalign{\smallskip}\hline
  \end{tabular*}
\label{contrib}
  \end{center}
\end{table*}

\section{Conclusion and Perspectives}

In spite of the new and precise data on the two-pion
spectral function from CMD-2 and SND, and on multihadron
cross sections from \babar , the lowest
order hadronic vacuum-polarization contribution remains the 
most critical component in the Standard Model prediction
of $a_\mu$. Yet, for the first time in recent years, the accuracy
of the prediction exceeds that of the experiment. One should not
forget however that the theoretical error is completely propagated
from the systematic uncertainties of the input experiments, the
estimation of which we totally depend. Also the evaluation of the
systematic uncertainty on the hadronic light-by-light contribution
is more subject to caution.

The discrepancy between the $2\pi$ spectral functions obtained from $\tau$
decays and from $e^+e^-$ annihilation is still unresolved: it affects
both the overall normalization and the shape. A particularly important
test of the relative normalization is obtained, when comparing the
measured branching fraction for $\tau^- \to \pi^- \pi^0 \nu_\tau$ to its
Standard Model prediction from the $e^+e^-$ spectral function, corrected
for small isospin-breaking effects. Here the discrepancy amounts to
4.5 standard deviations.

In view of this problem and the fact that both Novosibirsk experiments
agree at the 1\% level, the hadronic contribution is computed only with
$e^+e^-$ data, excluding for the moment KLOE until their disagreement
with Novosibirsk is understood. The choice of using only two experimental
inputs (CMD-2 and SND) among the four available (+KLOE and $\tau$ decays)
is clearly not satisfactory and even debatable, but represents in our view
the most reasonable solution, all arguments considered. In this scenario
we find that the Standard Model prediction of $a_\mu$ differs from the 
experimental value by 3.3 standard deviations.

We are looking forward to the forthcoming results on the 
two-pion spectral function from KLOE and \babar using the $\pi\pi/\mu\mu$
ratio. Since in this way vacuum polarization cancels, these data will 
help to reduce the systematic uncertainty due to the corrective 
treatment of radiative effects, always problematic in previous experiments
normalized by luminosity.
More data from \babar\ on multihadron final states are also expected soon.
With new experimental input to the vacuum polarization integrals, 
the quality of the prediction will improve, opening the way to 
a more precise direct determination of $a_\mu$~\cite{hertzog}. 
Unfortunately, some recent prospective work in the US~\cite{P5} 
does not cast a bright future in this direction.

%
%

\section*{Acknowledgements}

I would like to thank Andreas H\"ocker, Simon Eidelman and Zhiqing Zhang 
for our fruitful collaboration.

\end{document}